\documentclass[a4paper,11pt]{article}
\usepackage[latin1]{inputenc}
\usepackage[english]{babel}
\usepackage{graphicx}
\usepackage{psfrag}
\usepackage{ushort}
\usepackage{amsmath}
\title{A mechanical mode-stirred reverberation chamber with chaotic geometry}
\author{Gabriele Gradoni, \\ \small School of Mathematical Sciences, \\ \small University of Nottingham, \\ \small University Park - NG7 2RD \\ \small United Kingdom \\
Franco Moglie, Valter Mariani primiani \\ \small Dipartimento di Ingegneria dell'Informazione, \\ \small Universit\`a Politecnica delle Marche \\ \small Via Brecce Bianche - 60131 \\ \small Italy}
\date{July 6, 2014}
\begin{document}
\maketitle

\begin{abstract} 
A previous research on multivariate approach to the calculation of reverberation chamber 
correlation matrices is used to calculate the number of independent positions in a mode-stirred 
reverberation chamber. 
Anomalies and counterintuitive behavior are observed in terms of number of correlated matrix 
elements with respect to increasing frequency.
This is ascribed to the regular geometry forming the baseline cavity (screened room) of a reverberation chamber, 
responsible for localizing energy and preserving regular modes (bouncing ball modes). 
Smooth wall deformations are introduced in order to create underlying Lyapunov instability 
of rays and then destroy survived regular modes. Numerical full-wave simulations 
are performed for a reverberation chamber with corner hemispheres and 
(off-)center-wall spherical caps. Field sampling is performed by moving a mechanical 
``carousel'' stirrer. It is found that wave-chaos inspired baseline geometries improve 
chamber performances in terms of lowest usable frequencies and number of independent 
cavity realizations of mechanical stirrers. 
\end{abstract} 

\section{Performance of (or how chaotic are) reverberation chambers}
\begin{figure}[t]
\centering
\includegraphics[width=0.8 \textwidth]{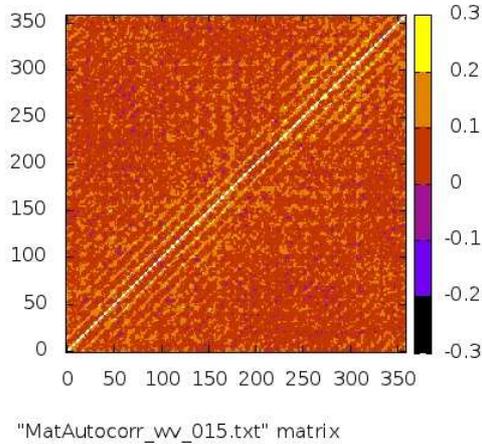}
\caption{\label{fig:rc_corr} An example of multivariate correlation function for the large Ancona RC facility at relatively low frequencies. A structured off-diagonal part is present.}
\end{figure}

\begin{figure}[t]
\centering
\includegraphics[width=0.8 \textwidth]{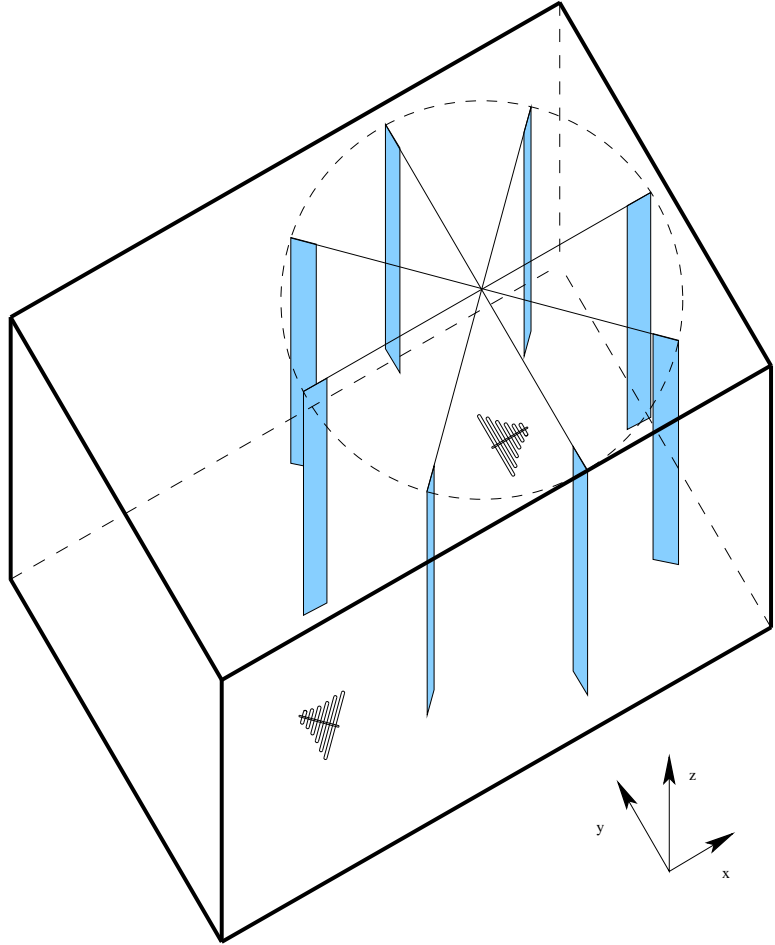}
\caption{\label{fig:rc_geo} Geometry of a the ``carousel stirrer'' as realized inside the small RC facility of the Laboratory of ElectroMagnetic Compatibility, Universit\`a Politecnica delle Marche, Ancona, Italy.}
\end{figure}

\begin{figure}[t]
\centering
\includegraphics[width=0.8 \textwidth]{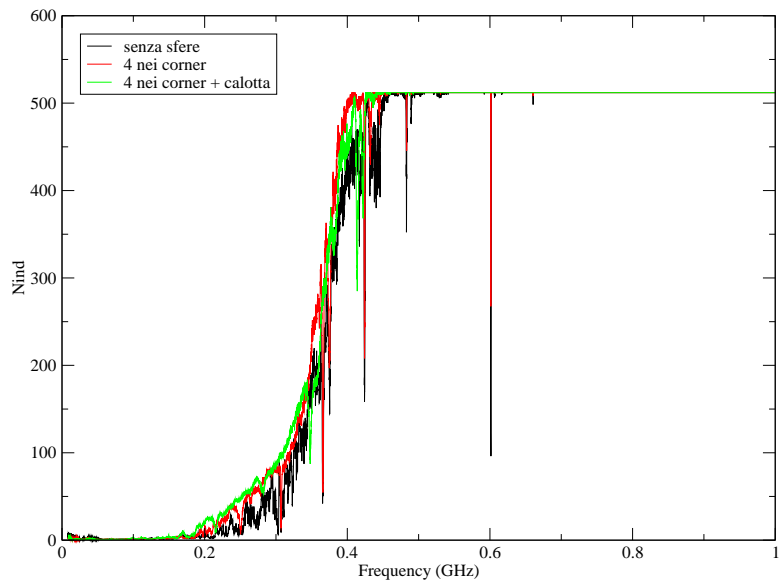}
\caption{\label{fig:Nind_Sph} Uncorrelated stirrer position versus frequency. The presence of corner spheres and spherical caps results beneficial in removing 
anomalies and in improving low frequency sirrer efficiency.}
\end{figure}

\begin{figure}[t]
\centering
\includegraphics[width=0.8 \textwidth]{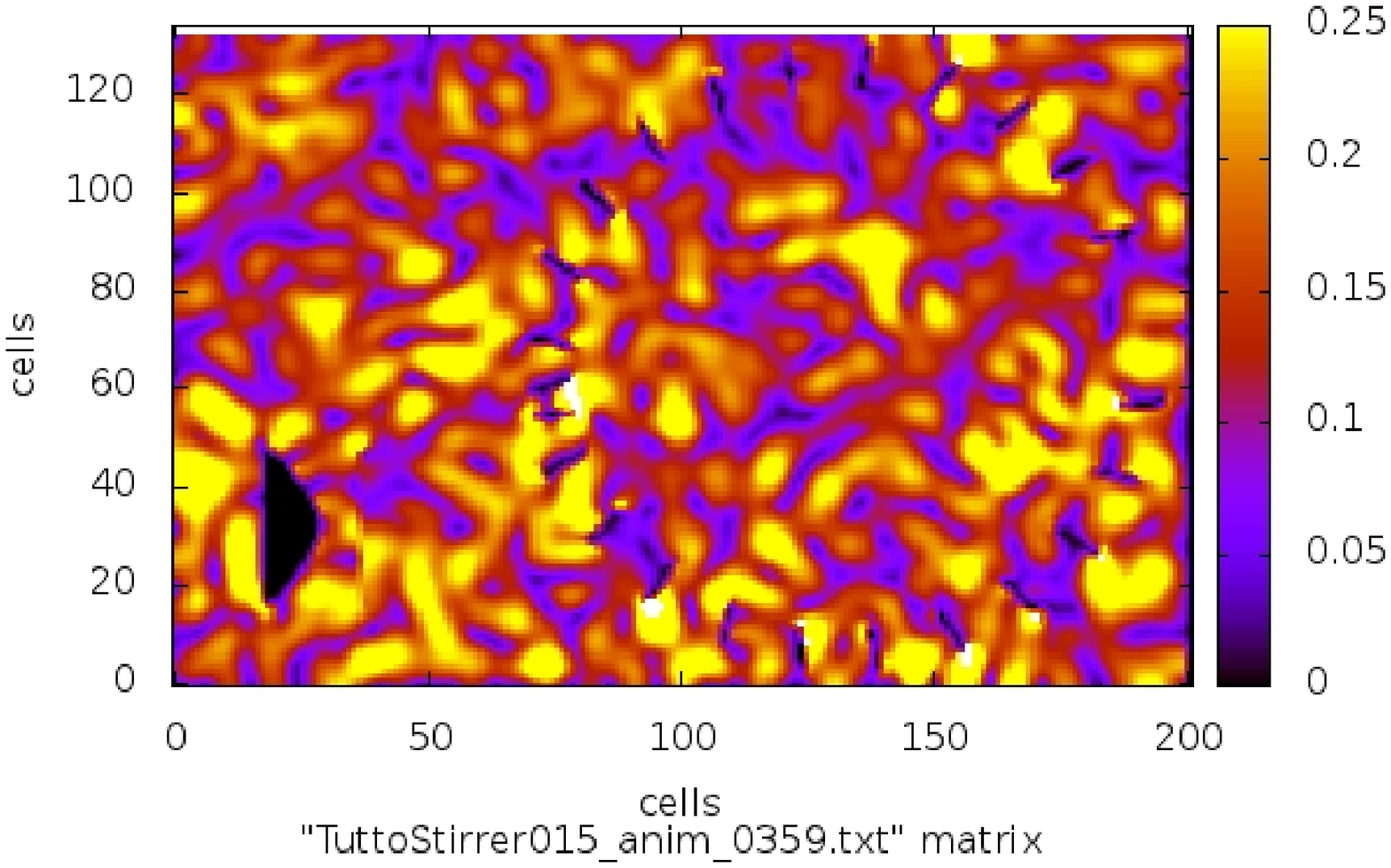}
\caption{\label{fig:field_2d} Chaotic (speckled) field distribution of a 2D plane of the 3D RC.}
\end{figure}

We started thinking seriously on how to revise the IEC strategy for calculating the number of uncorrelated stirrer positions back in 2010. 
In the past scientific literature, 
there exist other field-based strategies to measure the performance of mechanical stirrers \cite{Corona_RCres_2002,Marvin_RCperf_2005,Holloway_RCload_2006,Serra_WavImp_2014}.
Inherently, the structure of the field-field correlation matrix carries information about the perturbation 
process, e.g., mechanical \cite{Kostas1991}, frequency \cite{Hill1994}, space \cite{2001_Arnaut}, time, source \cite{Cerri_SourStir_2009}, 
or hybrid \cite{Gradoni_GEV_2010} stirring. 
An example of multivariate correlation matrix \cite{Corr_2013_Gradoni} having stirrer position pairs as elements, calculated at fixed frequency over a large chamber region, is reported in Fig. \ref{fig:rc_corr}.
The low-frequency RC behavior causes a structured off-diagonal part to appear. In the high-frequency regime, where 
\emph{spatial uncorrelation} takes over and leads to \emph{stir uncorrelation}, i.e., uncorrelation between stir trace pairs, the RC correlation matrix assumes a purely diagonal structure. 
Based on the concept of multi-variate correlation matrix, involving sampling through a large and dense spatial grid (lattice) 
rather than a single point measurement as suggested by the normative, we can estimate the number of uncorrelated stirrer positions 
more reliably. 
This actually allows for taking the spatial chaoticity of the RC into account along with the effect of moving the stirrer, as a whole.  
Finite-difference time-domain (FDTD) simulations highlighted an \emph{overestimation} of the traditional strategy. 
The multi-variate approach is believed to be more robust and reliable than the IEC definition as it naturally takes into account the chaotic distribution of fields through the 
sampling lattice. Such an approach can be employed to different physical quantity by keeping the ``thermodynamic'' equilibrium of chamber fields 
with respect to the (macroscopic) others. As an example, it is possible to calculate the number of independent lattice points , or 
the number of independent frequencies used in the sliding average (electronic stirring \cite{Hill1998}) w. r. t. frequency. 
The outcome of such an exploration is reported in \cite{Gradoni_indPos_2014}. 
In that paper, a different approach to relate the number of independent positions to the lowest usable frequency (LUF) has been proposed based 
on intra-distributions of correlation matrix elements. 
The EMC research community actually followed this new line of enquiry along with us 
through the last years \cite{RC_2012_Krauthauser,RC_2012_Chen,RC_2014_Monsef}.
The proliferation of independent stirrer positions increases linearly from the LUF, and reaches a maximum after a transition region that 
is in the order of hundreds of MHz in an actual RC. Anomalies have been discovered where this number drops down suddenly in very 
localized frequency narrow-bands, also belonging to the high frequency regime.  
This can be explained as follows: firstly the inefficiency of the irregular stirrer in breaking up cavity symmetry at \emph{any} position and \emph{any} frequency; 
secondly the transition from a position to another (arbitrary) one can leave residual regular modes through the detection (sub)volume. 
Consequently, two expedients can be adopted to improve an RC: 
\begin{itemize} 
 \item improve the mechanical stirrer geometry, also in view of its motion relatively to the baseline (room) cavity;
 \item change the baseline cavity geometry in order to create field disorder independently on stirrer geometry. 
\end{itemize}
The first strategy has been fulfilled through the proposal of a new stirrer geometry: the so-called ``carousel stirrer'' \cite{RC_2012_Moglie}.  
In this structured, equidistant $z$-folded paddles are distributed through the exterior surface of a cylinder, which rotates 
to create cavity realizations. A scheme of the Ancona small RC facility equipped with this stirrer is reported in Fig. \ref{fig:rc_geo}. 
The second strategy can be fulfilled through through geometries inspired by wave chaos, as pioneered in \cite{Gros2014664}. 
In particular, chaotic states have been observed in the high frequency limit of a 3D Sinai-like. 
However, scars have been found at relatively high frequencies, surviving wall deformations. This field enhancement localized 
at unstable orbits can originate from bouncing ball \cite{Stockmann2013} as well as tangential \cite{Gros2014664} modes.
An example of bouncing ball scar in 3D cavities is reported in \cite{Stockmann2013}.
To overcome this problem, in \cite{Gros2014664} it is suggested to use multiple hemispheres and spherical caps at the center or 
off the center of cavity walls in order to decrease the LUF of real-life reverberation chambers.  
This idea is pursued and refined in this study. 
\emph{In particular, we find that using an RC inspired by wave chaos geometries improves the performance, i.e., the effectiveness, of 
mechanical mode-stirrers.}
Figure \ref{fig:rc_geo} shows a few examples of baseline 
geometry simulated with an in-house FDTD code in order to reduce or even cancel anomalies 
in the production of uncorrelated stirrer positions. 
\begin{figure}[t]
\centering
\includegraphics[width=0.8 \textwidth]{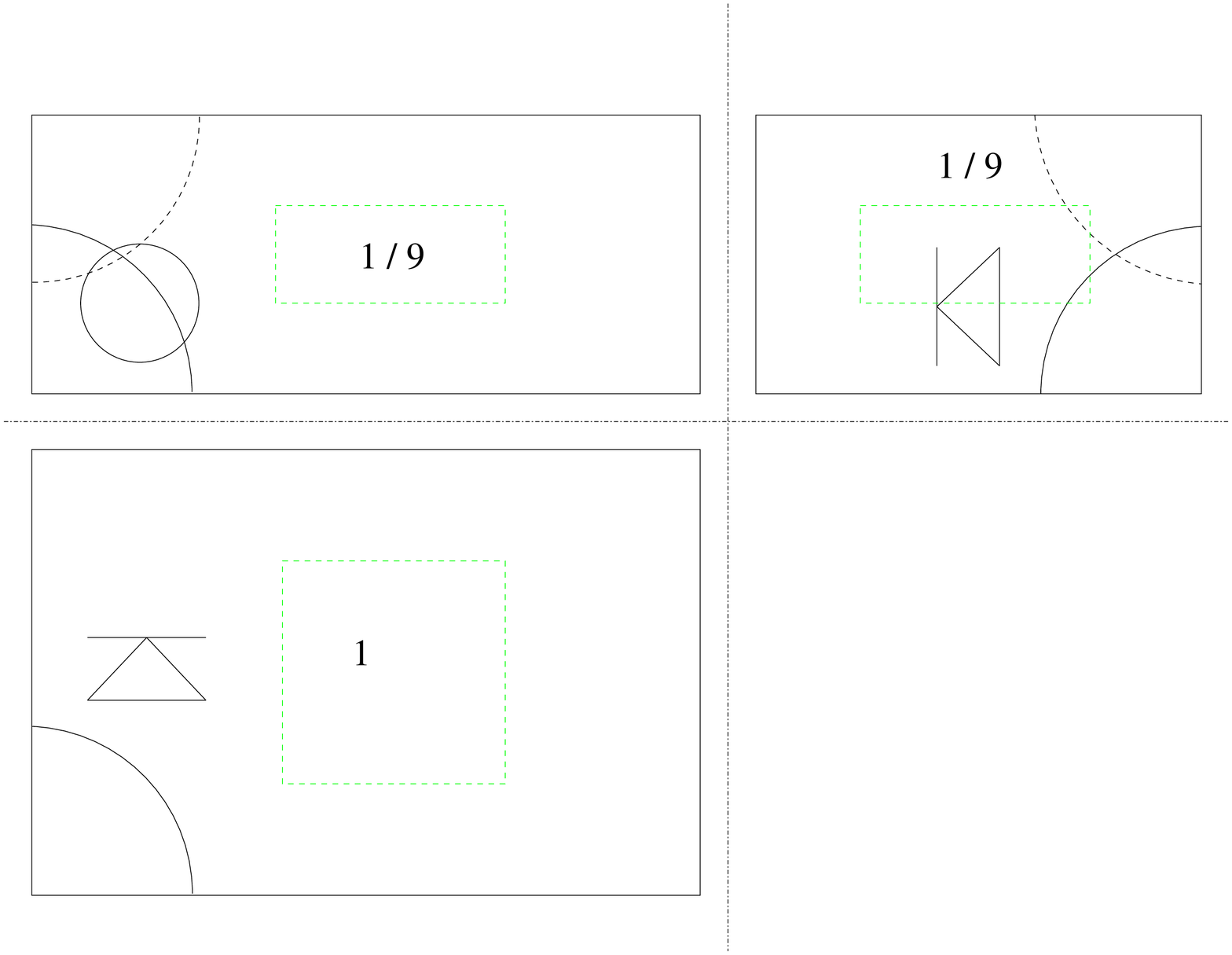}
\caption{\label{fig:rc_sfere} Geometry of a reverberation chamber with walls deformed through hemispheres and spherical caps.}
\end{figure}
Therefore, we believe the joint use of wall deformations and mechanical stirrers to be a good solution. 
Figure \ref{fig:Nind_Sph} shows the comparison between RC equipped with a ``carousel'' stirrer with and without 
deformations. 
It is worth noticing that, being cylindrical, the ``carousel'' geometry creates a (discrete, irregular) Sinai-like smooth  surface 
enhancing wave chaotic regime. This is confirmed by FDTD full-wave simulations 
of the RC. The resulting field distribution is reported in Fig. \ref{fig:field_2d} for a 2D plane of the 3D RC, which is clearly specked 
as for chaotic states. 

Besides reducing anomalies, there are other interesting consequences.
Interestingly, we notice the presence of much more ergodic modes at low frequencies as a consequence of the 
chaotic dynamics created by spherical surfaces. 

Wave chaos implies high sensitivity to the cavity boundaries. We expect that the number of independent positions of the stirrer might 
change for different boundary geometries, i.e., either the polygonal geometry proposed originally or the wave-chaos inspired geometry proposed recently 
should behave differently. 

Further work is in progress to visualize fields at those anomalous frequencies, where 
we expect to find scars and regular tangential modes that localize energy in spatial region not affected by the presence of a stirrer.

\section{Acknowledgements} 
G. G. acknowledges hospitality and financial support from DII, Universit\`a Politecnica delle Marche, Ancona, Italy, within the Campus World program ``Visiting Scientist'', 
through May 2014. 


\end{document}